\documentclass[twocolumn,showpacs,prb]{revtex4}

\usepackage{amssymb}
\usepackage{amsmath}
\usepackage{epsf}
\usepackage{graphicx}
\usepackage[dvips]{epsfig}
\usepackage{bm}

\begin{document}

\title{Absorption Cross-Section and Near-Field Enhancement in
Finite-Length Carbon Nanotubes in the Terahertz-to-Optical Range}

\author{M. V. Shuba\footnote{Author to whom correspondence should be addressed: Tel.: +37517 2264220; fax: +37517 2265124. \textit{E-mail address:} mikhail.shuba@gmail.com.}, S. A. Maksimenko and G. Ya. Slepyan, }
\affiliation{\\ Institute for Nuclear Problems, Belarus State
University, Bobruiskaya 11, 220050 Minsk, Belarus}

\begin{abstract}
Electromagnetic characteristics of single-walled finite-length
carbon nanotubes -- absorption cross-section and  field enhancement
in the near zone -- are theoretically studied in a wide frequency
range from terahertz to visible. The analysis is based on the
impedance-type effective boundary conditions  and the integral
equation technique. Comparison with experimental results is carried
out allowing qualitative physical interpretation of low-frequency
(far-IR and terahertz) absorption band observed in experiments.
Potentiality of CNTs for the IR photothermolysis of living cells is
discussed. Strong local field enhancement is predicted to be
inherent to metallic CNTs in the near-field zone providing necessary
mechanism for far-IR and terahertz near-field optics.

\textbf{Keywords}: Carbon Nanotube, Absorbtion Cross-Section,
Near-Field Enhancement.
\end{abstract}
\pacs{42.70.-a, 73.25.+i, 77.84.Lf, 78.67.Ch}
 \maketitle

\section{Introduction}

Promising potentiality of nano-scale objects and nano-structured
systems for transmission and processing electromagnetic signals
motivates active and permanently growing investments into studying
their electromagnetic response.  A new branch of physics of
nanostructures — nanoelectromagnetics — is currently emerging. Among
others, carbon nanotubes \cite{
Reich_b04} (CNTs) are of special interest owing their unusual
electronic and mechanical properties, temperature stability and
quasi-one-dimensional nature. Single-walled CNTs are  hexagonal
networks of carbon atom rolled up into cylinder $\sim 1-3\,$nm in
diameter and $\sim 0.1-10\, \mu$m in length. Recently, CNTs were
proposed for realization of different integrated circuits elements
and electromagnetic devices, such as transmission lines
\cite{Slepyan99,Maksim00,Hagmann_05,Rybczynski_07_APL},
interconnectors
\cite{Raychowdhury_06,Miano_06,Chiariello_07,Maffucci_08,Li_08} and
nanoantennas
\cite{Wang_04_APL,Hanson05,Slepyan06,Burke06,Hanson06,Kempa_07,Shuba07,Nemilentsau,Wang_08_IJIRMW,Maksimenko07,Hanson08}.
First experimental realization of the CNT-based radio have been
reported \cite{Rutherglen,Iensen} and potentiality of CNTs as
terahertz and infrared light emitters has been demonstrated
\cite{Misewich_03_Sci,Chen_05_Sci,Kibis3,Kibis_07_NL,Batrakov_06_SPIE,Kuzhir_07_SRIMN,Batrakov_08_PhE}.
Strong absorbance of CNTs in optical  and terahertz ranges
\cite{O'Connel_02,Bommeli,Ugawa,Zhao,Kang} can be used for
electromagnetic stimulation of nanotubes inside living cells to
afford various useful functions\cite{Kam_05} and, in particular, for
the selective cancer
photothermolysis\cite{Kam_05,Panchapakesan_05,Gannon_07,Yang_07}.
Note that the plasmonic photothermal therapy using gold
nanoparticles has emerged to be highly promising for cancer therapy
and is fast developing \cite{Zharov_05,Lapotko_2007,Huang_07}.

Obviously, the nanophotothermal effect is expected to be efficient
only in the maximal-absorption band whether it be the plasmon
resonance in metal nanoparticles or geometrical (antenna)
resonance\cite{Slepyan06} in CNTs. As different from the plasmon,
the antenna operating frequency is dictated by the CNT length and
therefore provides spectral tunability of the method. Moreover, the
strong slowing-down effect characteristic for the surface wave
propagation in CNTs\cite{Slepyan99} shifts geometrical resonances to
the red opening far-infrared and terahertz ranges for the
nanothermolysis. Although water is opaque in the terahertz range
preventing most living cells from the CNT-based terahertz
photothermal therapy, the method can be useful for tissue with low
water content, e.g. fatty tissue. Of course, elaboration of a
consistent theory of the CNT-based photothermolysis of living cells
is a global scientific  problem comprising many physical, chemical
and biological subtasks. Among others, the heat transfer between CNT
and fluidic cell volume is of special interest, first of all because
it is governed by specific thermodynamics of CNTs. It should be
emphasized that the nanothermodynamics as a
whole\cite{Gemmer_b05,Volz_b07} and, in particular, the
thermodynamics of CNTs\cite{Berber_00,Che_00,Kuroda_05,Pop_06}
essentially differ from the classical  macroscopic thermodynamics.
However, even so, the absorbtion cross-section  $\Lambda(\lambda)$,
where $\lambda$ is the wavelength, remains to be a basic quantity
defining exterior heat source in the cell.

That is why theoretical prediction of spectral properties of the
CNT absorbtion cross-section $\Lambda(\lambda)$ is a critical
problem for many medical and electromagnetic applications of CNTs.
In our paper we present a theory of electromagnetic wave
absorption by finite-length carbon nanotubes in a wide spectral
range from terahertz to visible frequencies. The theory allow
correct interpretation of experimental results reported in
Refs.\cite{Bommeli,Ugawa,Zhao,Kang} -- a resonant behavior of
dielectric properties of single-walled CNT film in terahertz and
infrared regimes. Explanations of the behavior by the phonon
resonance\cite{Kang,Dresselhaus02} or energy gap due to the CNT
curvature\cite{Kane} do not respond the question properly. Indeed,
as was mentioned by Bommeli \emph{et al.}\cite{Bommeli},  the
far-infrared (terahertz) absorption resonance is temperature
independent and therefore can not be related to phonon modes.
Moreover, the absorption peak shifts in frequency when measuring
different specimens, and consequently cannot be strictly
considered as an intrinsic properties of the nanotubes like a
phonon mode. This is reasonable argumentation against the phonon
model. As for the second explanation, no direct calculations of
the curvature impact on the scattering and absorbtion properties
of CNTs have been provided for comparison with experimental data.

Recently, we have proposed\cite{Slepyan06,Shuba07} alternative
explanation of the quasi-resonant behavior of dielectric properties
of CNTs in terahertz and far-infrared region. We have found that
owing to the strong slowing down of surface waves (plasmon-polariton
modes) in CNTs \cite{Slepyan99}, the geometrical (antenna)
resonances are shifted to the red and, for micrometer-length CNTs,
appear to be in the terahertz range. This is pure finite-length
electromagnetic effect which dictates the peculiarities of IR and
teragertz properties of CNTs. In the present paper we utilize the
finite-length CNT model combining the quantum-mechanical dynamic
conductivity \cite{Slepyan99} and the integral equations method
\cite{Ilyinsky}, for solving the electrodynamic boundary-value
problem. That allows us to give a satisfactory explanation of
experimental date \cite{Zhao} for the CNT film absorbance in a wide
frequency range. We also investigate the CNT ability to produce
highly localized electromagnetic fields -- extremely interesting
property for near-field optics, realization of the Parcell effect
for terahertz-range emitters and as a means for the surface plasmon
excitation in metallic substrates.

\section{Theoretical consideration}

In theoretical analysis of electromagnetic properties of
finite-length CNT we combine methods of classical electrodynamics
and semiclassical physical kinetics, see
Refs.\cite{Slepyan99,Slepyan06}. The latter means that the motion of
$\pi $-electrons in CNT is described as classical motion of
quasi-particles with quantum dispersion law accounting for the
hexagonal crystalline structure of graphene and quantization of the
transverse momentum. This allows us to formulate effective boundary
conditions for electromagnetic field on the CNT surface in the form
of two-side anisotropic impedance boundary conditions
\cite{Slepyan99}. The CNT electronic properties are incorporated
into analysis by means of the surface impedance tensor. In this case
the problem of the CNT-based antenna is reduced to the
boundary-value problem of classical electrodynamics.

Let an isolated single-walled CNT of length $L$ and cross sectional
radius $R$ be aligned parallel to the $z$ axis of the cylindrical
coordinate system ($\rho,\varphi,z)$. The origin of coordinate
system is located at the point $z = 0$ in the geometrical center of
the CNT. The nanotube is exposed to external field with
$E_z^{(0)}(z,R)\exp (-i\omega t)$ as  $z$-component; $\omega$ is the
angular frequency. This field induces in CNT axial surface current
of the density $\mathbf{j}(\mathbf{r})$, which reradiates the
scattered field. Assuming the CNT radius to be small as compared to
the free-space wavelength, we neglect the transverse current in CNT.
We also neglect azimuthal variations of the axial current on the CNT
surface, that is we set $\mathbf{j}(\mathbf{r}) = j(z){{\bf e}}_z$,
where ${\bf e}_z$  is the unit vector along the CNT axis.

The electric Hertz potential of scattered field $\Pi (\rho,z)$
satisfies the Helmholtz equation and radiation conditions, and can
be represented in the form of single-layer potential
\cite{Slepyan06}:
\begin{equation}
\label{eq1} \Pi (\rho,z ) = \frac{iR}{\omega }\int\limits_{ - L /
2}^{L / 2} {j ({z}')G (z - {z}',\rho ,R )d{z}'} \,,
\end{equation}
\noindent where
\begin{equation}
\label{eq2} G(z,\rho ,R) = \int\limits_{0}^{2\pi} {\frac{\exp
\left( {ik\sqrt {\rho ^2 + R^2 - 2R\rho \cos \varphi + z^2} }
\right)}{\sqrt {\rho ^2 + R^2 - 2R\rho \cos \varphi + z^2}
}d\varphi }\,.
\end{equation}
Imposing boundary conditions on the  CNT surface\cite{Slepyan99} we
arrive at the integral equation for the current density:
\begin{eqnarray}
\int\limits_{ - L / 2}^{L / 2} E^{(0)}_{z} ({R,z}')e^{ik\vert z -
{z}'\vert }\,d{z}' + C e^{ikz} + D e^{ - ikz}\qquad\qquad
    \cr \rule{0in}{5ex}
    =\! \int\limits_{ - L /
2}^{L / 2}\!  \left[ {\frac{4\pi  R }{c }G(z - {z}',R ,R )}
 +  {\frac{1}{\sigma }e^{ik\vert z - {z}'\vert }} \right]
 j({z}')\,d{z}'\,,\,\label{eq3}
\end{eqnarray}
where $\sigma$ is the CNT axial conductivity derived via quantum
transport theory \cite{Slepyan99}, whereas $C$ and $D$ are unknown
constants to be determined from the edge conditions
\begin{equation}
\label{eq5} j(\pm L / 2) = 0,
\end{equation}
which express the absence of concentrated charges on
the CNT edges $z$= $\pm L$/2.

Generally, Eq. (\ref{eq3})  can not be solved analytically. For
numerical solution the integral on the right side of (\ref{eq3}) is
numerically handled by a quadrature formula, thereby transforming
Eq. (\ref{eq3}) into a matrix equation. Solution of matrix equation
gives axial current density $j(z)$ along CNT. The spatial
distribution of the non-zero component of the scattered
electromagnetic field in arbitrary point can be found as follows:
\begin{eqnarray}
\label{eq6} 
E_{\rho} = \frac{\partial ^2\Pi}{\partial
\rho\partial z}\,,
~~    E_z =\Bigl( {\frac{\partial ^2}{\partial
    z^2} + k^2} \Bigr)\Pi \,,
~~H_{\phi} = ik\frac{\partial \Pi }{\partial \rho }\,.
\end{eqnarray}

The CNT absorption cross-section along the $z$-axis is determined by
the relation
\begin{equation}
\label{eq7} \Lambda = P_t / I_0\, ,
\end{equation}
where
\begin{equation}
\label{eq8} P_t = \pi R {\rm Re}\left( \frac{1}{\sigma } \right)
\int_{ - L / 2}^{L / 2} {\vert j (z)\vert ^2dz}
 \end{equation}
is the power loss due to current dissipation, and $I_0 $ is the
intensity of incident electromagnetic wave; for plane wave $I_0 = (c
/ 8\pi )\vert E^{(0)}_z\vert ^2$.

Electromagnetic characteristics of CNT demonstrate qualitatively
different behavior in two fundamentally distinguishing regimes.
The first one, further referred to as the Drude conductivity
regime, is characterized by the propagation of low-attenuated
surface waves \cite{Slepyan99} and corresponding geometrical
resonances \cite{Slepyan06} in finite-length CNTs. The resonances
are due to intraband motion of conducting electrons. The second
one -- called as the optical transitions  regime -- is determined
by the interband transitions of electrons. In the Drude
conductivity regime, CNT is analogous  in many respects to
macroscopic RF wire antenna
\cite{Hanson05,Slepyan06,Burke06,Miano_06}. Regime of optical
transitions has quantum nature and, consequently, has no
macroscopic analogs. The angular frequency $\omega_e $, which
divides the Drude conductivity regime ($\omega < \omega_e )$ and
the regime of optical transition ($\omega > \omega_e )$, depends
on the electronic and geometric properties of concrete CNT. From
the approximate relation for the density of electron
states\cite{Mint_98} one can found that $\omega_e \approx
2\upsilon _F / R$ and $\omega_e \approx 2\upsilon _F / (3 R)$ for
metallic and semiconducting CNT, respectively; $\upsilon _F$ is is
the $\pi$-electron velocity at the Fermi level.

As has been shown in Refs.\cite{Slepyan06,Hanson06}, in the range of
interband transitions the surface waves in single-walled CNTs are
strongly attenuated. Therefore, the surface current density $j(z)$
in a nanotube exposed to external electric field obeys with high
accuracy the Ohm's law,
\begin{equation}
\label{eq9} j(z) = \sigma E_z^{(0)} (z)\,,
\end{equation}
which is indeed the first Born approximation of scattering theory
in application to Eq. \eqref{eq3}.

The use of Eq. (\ref{eq9}) as an approximate solution of the
integral equation (\ref{eq3}) is only possible when the local
electric field on the CNT surface is much smaller than the external
electric field, i.e. the relation
\begin{equation}
\label{eq10} \left| {\left( {\frac{\partial ^2}{\partial z^2} +
k^2} \right)\Pi (R,z)} \right| \ll \vert E_z^{(0)} (R,z)\vert
\end{equation}
holds true over the CNT length.

Note that the solution (\ref{eq9}) does not satisfy the edge
conditions (\ref{eq5}). However, the error is strongly localized in
the vicinity of the edges and therefore does not influence the field
formation in far-field region; analogous situation appears, for
example, in the theory of diffraction by an aperture in infinitely
thin perfect screen \cite{Ilyinsky}. A comparison of the exact
solution by Eq. (\ref{eq3}) with approximate calculations by Eq.
(\ref{eq9}) shows that the latter one can serve for high-accuracy
evaluation of CNT scattering and absorption parameters in the
interband transition regime.

Substitution of  (\ref{eq9}) into Eqs. (\ref{eq8}) and (\ref{eq7})
leads to the simple formula for the absorption cross-section of
isolated CNT in the interband transitions regime in the
$z$-direction:
\begin{equation}
\label{eq11} \Lambda = \frac{8\pi ^2RL}{c}\,{\rm Re}( \sigma).
\end{equation}
In that regime, this formula  can be directly applied to calculation
of the absorbance of a bundle of parallel CNTs. As has been shown by
Hao and Hanson \cite{Hanson06}, in the regime of interband
transitions the electromagnetic coupling of carbon nanotubes
composed in a planar array is very low. Therefore, the current in an
individual tube of the array can be found from (\ref{eq9}).
Extending that result to CNT bundle comprising $N$ nanotubes, we
arrive at the formula
\begin{equation}
\label{eq12} \Lambda _c = \sum\limits_{m = 1}^N {\Lambda _m }\,
\end{equation}
for the absorption cross-section of the bundle in the direction
parallel to its axis. In this expression, $\Lambda _m $ is
absorption cross-section of $m$th CNT. Because of the strong
electromagnetic coupling of metallic CNTs in the Drude conductivity
regime \cite{Shuba07}, in this regime the formula (\ref{eq12}) can
not be applied to bundles comprising more than one metallic CNT.

\section{Numerical results}

\subsection{CNT absorption cross-section}

\begin{figure}[!htb]
\begin{center}
\includegraphics[width=3.1in]{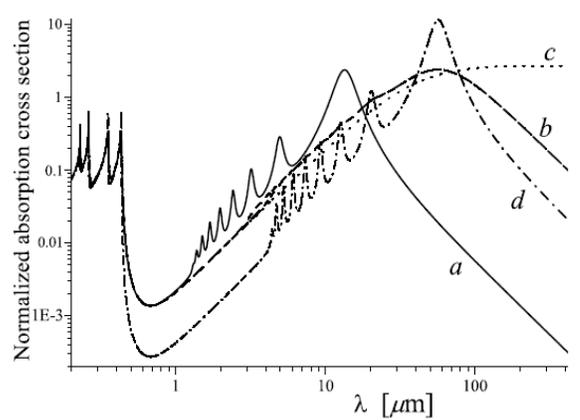}
\end{center}
\caption{Normalized absorption cross-section $\Gamma(\lambda)$ of
(9,0) zigzag CNT; \emph{curve a:} $L = 100\,$nm, $\tau = 2\times
10^{-14}\,$s, \emph{curve b:} $L = 500\,$nm, $\tau = 2\times
10^{-14}\,$s, \emph{curve c:} $L = 6000\,$nm, $\tau = 2\times
10^{-14}\,$s, \emph{curve d:} $L = 500\,$nm, $\tau = 1\times
10^{-13}\,$s.} \label{fig1}
\end{figure}

Figure \ref{fig1} demonstrates the normalized absorption
cross-section $\Gamma(\lambda)=\Lambda(\lambda)/(2RL)$  of (9,0)
metallic zigzag CNT for different CNT lengths $L$ and different
electron mean free-path time $\tau$. The value $\tau$ is used under
calculation of the axial conductivity\cite{Slepyan99} $\sigma$ and
is assumed to be constant over the whole frequency range considered.
As it is seen,  three different bands are distinguished in Fig.
\ref{fig1}, characterizing the (9,0) CNT interaction with
electromagnetic field: the Drude conductivity regime ($\lambda
\gtrsim 1\,\mu$m), the optical transitions regime ($\lambda
\lesssim 0.5$ $\mu$m), and the intermediate regime ($0.5< \lambda
< 1 \,\mu$m).

In the regime of Drude conductivity the cross-section
$\Gamma(\lambda)$ demonstrates a set of resonances. Intensity of
resonances grows smaller as  wavelength decreases, whereas the
length  $L$ increase shifts the resonances to the red and leads to
their broadening (compare lines $a$ and $b$ in Fig. \ref{fig1}),
so that they completely disappear at sufficiently large $L$ (see
line $c$). The same effect is observed with the increase of
$\tau$, compare lines $b$ and $d$, while the resonant wavelengths
do not depend on $\tau$. Physically, the resonances depicted in
Fig. \ref{fig1} are the geometrical (antenna)
resonances\cite{Slepyan06} whose wavelengths are dictated by the
condition of the standing surface wave to be settled. In view of
conditions (\ref{eq5}) the resonant wavelengths are approximately
determined by
\begin{equation}
\label{eq13} 2Lc = s\lambda_s v(\lambda_s) , \quad s = 1,3,5...\,,
\end{equation}
where $v (\lambda)$ is the surface wave phase velocity  at the
wavelength $\lambda$. It should be noted, that in the Drude
conductivity regime the surface waves in CNT are strongly slowed
down\cite{Slepyan99}: $v (\lambda) \ll c$. Therefore, accordingly
to (\ref{eq13}) the resonant wavelength is much longer than the
CNT length: $\lambda_s \gg L$. For example,  line $a$ in Fig.
\ref{fig1} shows three first geometrical resonances of 100 nm
length CNT at $\lambda_1 = 12.7\,\mu$m, $\lambda_2 = 4.9\,\mu$m
and $\lambda_3 = 3.2\,\mu$m.

In the short-wavelength regime ($\lambda < 0.5$ $\mu$m) the spectral
dependence $\Gamma (\lambda)$ also demonstrate  a set of resonances,
which are due to $\pi$-electron  transitions between valence and
conduction bands. Obviously, in this regime the normalized
absorption cross-section does not depend on the CNT length (see Fig.
\ref{fig1}) and is completely determined by the CNT conductivity
accordingly to Eq. (\ref{eq11}). The CNT conductivity resonances
correspond\cite{Slepyan99} to Van-Hove singularities of the density
of states of $\pi$-electrons.

In the intermediate region, both interband and intraband motion of
$\pi$-electrons contribute into the CNT conductivity. Our
calculations shows that in this range the quantity $\Gamma
(\lambda)$ turns out to be very small, has no resonances and
practically does not depend on the CNT length (compare lines $a$,
$b$ and $c$ in Fig. \ref{fig1}), while demonstrates strong
dependence on $\tau$.
\begin{figure}[!htb]
\begin{center}
\includegraphics[width=2.9in]{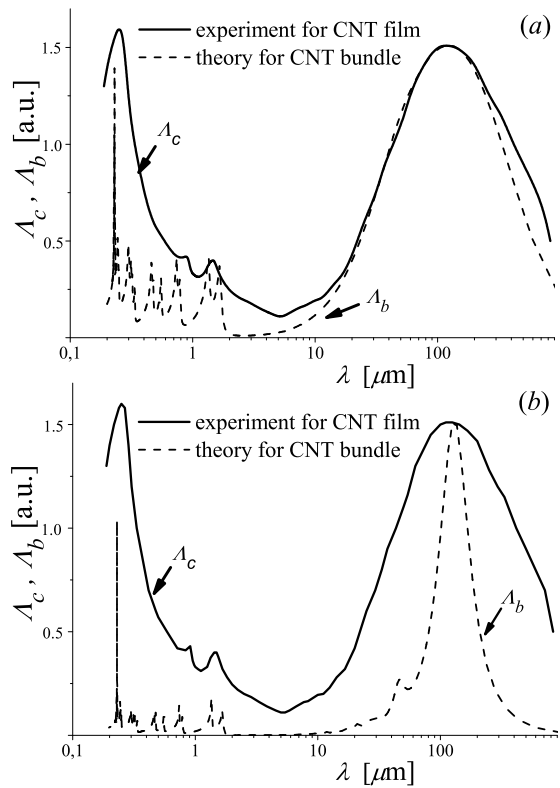}
\end{center}
\caption{Comparison between experimentally observed  normalized
absorbance $\Lambda_c$ of
single-walled CNT film, taken from Fig. 6 in Ref. \cite{Zhao},   and calculated normalized
absorption cross-section $\Lambda_b$ of CNT-bundle.
The 1.2 $\mu$m length  CNT bundle consists of three zigzag tubes
with chiral vectors (13,0), (12,0) and (11,0), respectively. In
calculations we  set $\tau = 2.2 \times 10^{ - 14}$ s  for plot
$(a)$ and $\tau = 1 \times 10^{ - 13}$ s for plot $(b)$.}
\label{fig2}
\end{figure}

\subsection{Comparison with experiment}

The experimentally registered by Hu \emph{et al.} \cite{Zhao}
normalized absorbance $\Lambda _c$ of a film comprising a sparse
disordered array of finite-length bundles of single-walled CNTs is
depicted in Fig. \ref{fig2} by solid line. The plot has been
extracted from Fig. 6 of cited article. In this article, the CNT
bundles were found to have average diameter 2.7 nm and length $L <
2$ $\mu$m. The dashed curves in Fig. \ref{fig2} show calculated
absorption cross-section $\Lambda _b$ of a bundle of three zigzag
tubes (one metallic and two semiconductor CNTs) with chiral
vectors (13,0), (12,0) and (11,0). The semiconductor nanotubes
(13,0) and (11,0) have been chosen because their first interband
transitions fit well the experimentally observed extremums in the
high-frequency part of the absorption spectrum, as it is seen in
Fig. \ref{fig2}. The low-frequency pick on the theoretical curve
is completely due to geometrical resonance of surface wave in
metallic (12,0) CNT, whereas the contribution of semiconductor
CNTs in this frequency range is negligibly small. Position of the
peak on the frequency axis is determined by the CNT length, while
its width is dictated by the electron free-path time $\tau$. In
figure \ref{fig2}(a) both parameters have been chosen to provide
the best correlation with the experimental plot: $L=1.2\,\mu$m and
$\tau = 2.2 \times 10^{ - 14}$ s. The later parameter turned out
to be very close to that in graphite.

As one can see, theoretical curve follows the main peculiarities
of the experimental one. Of course, variation of number of CNTs in
the bundle and their chiral vectors modifies positions of peaks
and their intensities. However, overall picture of the phenomenon
is kept unchanged permitting the use of the bundle absorbtion
cross-section as a model of the absorption in composite film with
CNT bundles embedded. The model allows a qualitative physical
interpretation of experimental results. In particular, as
conductivity of all CNTs has plasmon resonance in the ultraviolet
range at $\lambda _{\rm pl} = c\pi \hbar / \gamma _0=230$ nm (for
$\gamma _0 = 2.7$ eV), the dependence $\Lambda_b (\lambda)$ also
has a resonance at this wavelength. This resonance is well defined
in the experimental plot. Thus, the main conclusion which follows
from the comparison  is that the experimentally observed
absorption peak\cite{Bommeli,Ugawa,Zhao,Kang}, laying below the
range of interband transitions,  can certainly be attributed to
geometrical (antenna) resonances of constituting finite-length
CNTs, inhomogeneously broadened due to size dispersion in
composites. It should be emphasized that the problem of
inhomogeneous broadening is critical for the correct quantitative
interpretation  of absorption experiments in realistic CNT-based
composites  and requires distinct serious analysis which is far
beyond the scope of given paper.

Figure \ref{fig2}(b) shows the same absorbance of the composite
film as in Fig. \ref{fig2}(a) (solid line) and normalized
absorption cross-section of the same CNT bundle as in Fig.
\ref{fig2}(a) but for the value $\tau = 1 \times 10^{ - 13}$ s
(dashed line). The picture illustrates strong dependence of the
antenna peak linewidth of the CNT bundle on $\tau$.

\subsection {Near-zone field enhancement in finite-length CNT}

Currently, there exists a considerable interest to optics of metal
nanoparticles, largely due to their plasmonic properties
\cite{Novotny} and ability to produce giant and highly localized
electromagnetic fields \cite{Crozier,Kappeler}. Important
applications include microscopy \cite{Krug}, spectroscopy
\cite{Hillenbrand}, optoelectronic devices \cite{Puscasu} and, as
aforementioned, photothermolysis of living cells
\cite{Zharov_05,Lapotko_2007,Huang_07}. Naturally, one can expect
manifestation of analogous effects in metallic CNTs.  Since their
conductivity in the infrared (terahertz) regime has Drude-like
behavior, the propagation of surface waves along the CNT axis
\cite{Slepyan99} is provided, which are analogous to
plasmon-polariton wave in elongated metallic particles
\cite{Novotny}. The surface wave propagation is accompanied by the
field localization near the CNT edges. Further we discuss this
effect on more details.

The spatial structure of electric field in the near zone  is
conveniently characterized by the intensity enhancement factor
$\xi({\rm {\bf r}},\omega) = |{\rm {\bf E}} ({\rm {\bf
r}},\omega)|^2/|{\rm {\bf E}}^{(0)}|^2$, where $|{\rm {\bf E}}({\rm
{\bf r}},\omega)|^2$ is electric field intensity distribution and
$|{\rm {\bf E}}^{(0)}|^2$ is electric field intensity of incident
plane-wave illumination. One can expect that in the vicinity of
geometrical (antenna) resonance the spatial and frequency variables
are separated to a high accuracy allowing the enhancement factor
expression as
\begin{equation}
\label{eq15} \xi({\rm {\bf r}},\omega) \sim \frac{\varphi({\rm
{\bf r}})}{(\omega-\omega_1)^2+(\omega_1/2Q)^2} \,,
\end{equation}
where $\omega_1$ is the angular frequency corresponding to the
first geometrical resonance, $Q$ is $Q$-factor of the resonant
mode, and $\varphi({\rm {\bf r}})$ is a spatial distribution
function.

\begin{figure}[!htb]
\begin{center}
\includegraphics[width=2.9in]{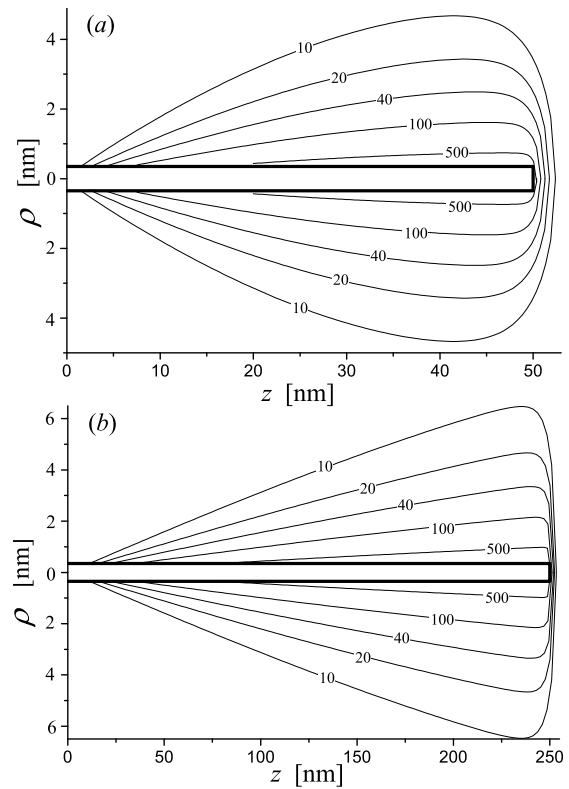}
\end{center}
\caption{The constant-value lines of the intensity enhancement
factor $\xi({\rm {\bf r}},\omega_1)$  in the vicinity of the right
half of (9,0) zigzag CNT with the radius $R=0.36$ nm and the
length (a) $L = 100$ nm, and (b) $L = 500$ nm. In both cases $\tau
= 2 \times 10^{ - 14}$ s. CNT is shown by the thick line. The
incident wavelength corresponds to the first antenna resonance:
(a) $\lambda _1 = 12.7$ $\mu$m and  (b) $\lambda _1 = 57$ $\mu$m.}
\label{fig3}
\end{figure}
The field intensity distribution near finite-length CNTs
illuminated by a plane wave with the electric field vector
directed along CNT axis is presented in Fig. \ref{fig3}. The
incident wavelengths correspond to the first geometrical
resonances of chosen CNTs. The constant-value lines of the
intencity enhancement factor $\xi({\rm {\bf
r}},\omega_1)\sim\varphi({\rm {\bf r}})$ are depicted. As the
field distribution at resonance frequency is symmetrical with
respect to the plane $z=0$, i.e. $\xi({\rm {\bf
r}},\omega_1)=\xi({\rm {-\bf r}},\omega_1)$,  the right half of
CNTs is only shown.

Figure \ref{fig3} demonstrates considerable, $\xi \sim 500$, and
increasing nearby edges  the near-zone field enhancement.  Such a
spatial distribution of the field intensity is dictated by general
principles of electrodynamics and can easily be understood from the
absence of free charges at the CNT tips (mathematically they are
geometrical singularities of the surface) resulting in the field
localization near the tips \cite{Ilyinsky}. The simplest example is
a perfectly conducting semiinfinite plane described by the spatial
distribution function $\varphi({\rm {\bf r}})\sim |{\rm {\bf
r}}|^{-1}$; here ${\rm {\bf r}}$ is the distance from the plane
verge. In more realistic models the verge is described by strongly
curved but regular functions\cite{Ilyinsky}.

\begin{figure}[!htb]
\begin{center}
\includegraphics[width=3.1in]{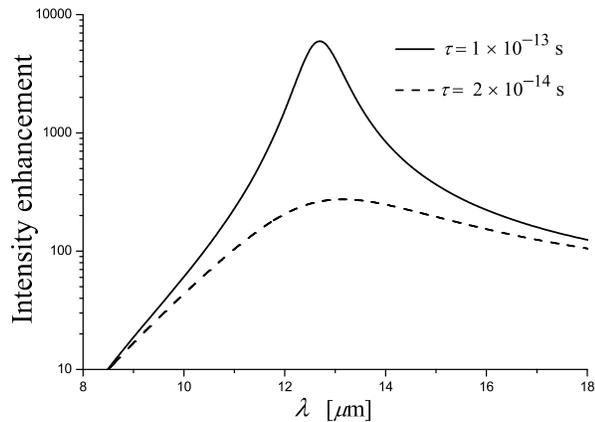}
\end{center}
\caption{The intensity enhancement factor $\xi({\rm {\bf
r}}_1,\lambda)$ vs wavelength for 100 nm length metallic zigzag
(9,0) CNT at different magnitudes of the free-pass time: $\tau =
1\times 10^{ - 13}$ s (solid line), and $\tau = 2 \times 10^{ - 14}$
s (dashed line). Calculation has been performed for the space point
given by cylindrical coordinates $\rho_1=1$ nm and $z_1=47.5$ nm of
the coordinate system presented in Fig \ref{fig3}(a).} \label{fig4}
\end{figure}

The intensity enhancement factor  versus wavelength calculated in
a fixed point ${\bf r}_1$ at different $\tau$ is depicted in Fig.
\ref{fig4}. As one can see, maximum magnitude of $\xi$ strongly
increases with $\tau$ and reaches e.g. 5000 for $\tau = 1 \times
10^{ - 13}$ s, which corresponds to the electron mean free path
$\sim 100$ nm. Experimental results indicate that  $\tau$ in
single-walled CNTs in the regime of Drude conductivity can be much
higher than in normal metal\cite{Hilt}. Thus, the intensity
enhancement factor 5000 and even bigger is quite reachable.

It can easily be shown that in the vicinity of the antenna
resonance $\lambda_1=2\pi c/\omega_1$ the dependence $\xi({\bf
r}_1,\lambda)$ depicted in Fig. \ref{fig4} follows the Lorentz
resonant line, supporting therefore our assumption resulted in Eq.
(\ref{eq15}). In the vicinity of resonance the enhancement factor
is reduced to a product of two partial independent factors, one of
which is completely determined by the  geometrical singularities
of the CNT at its tips, while the second one -- by the frequency
singularity due to antenna resonance. Note that the same structure
of the field enhancement factor, $L_\mathrm{pl}=L_{LR}L_{SPPR}$,
has been revealed for rough surfaces of metals\cite{Boyd_84} and
for metallic nanoparticles\cite{Crozier,Kappeler} in the vicinity
of plasmon resonance (the field and the intensity enhancement
factors, $L_\mathrm{pl}$ and $\xi$ respectively, are related by
$L_\mathrm{pl}=\sqrt{\xi}\,$). $L_{LR}$ and $L_{SPPR}$ are the
partial factors denoting to the lightning rod
effect\cite{Boyd_84,Kappeler} and plasmon resonance,
correspondingly. Comparison shows that the lightning rod effect
physically is identical to mechanism described for CNTs by the
factor $\varphi ({\bf r})$ ($L_{LR}=\sqrt{\varphi ({\bf r})}\,$),
whereas resonant coefficient in (\ref{eq15}) corresponds to the
factor $L_{SPPR}$: instead plasmons characteristic for metal
particles, surface waves propagate in CNTs. The difference is that
the field enhancement in metal particles is observed in the range
of plasmon resonance, i.e. in the wavelength range $400-1300$ nm,
while the CNT antenna resonance at realistic CNT lengths occupied
the far-IR and terahertz regions. The use of CNT bundles instead
isolated CNTs shifts the operating range in the  short-wavelength
direction \cite{Shuba07}.

\section {Conclusion}

In the paper, we have modeled absorption cross-section and
near-zone field enhancement in isolated single-wall carbon
nanotubes in a wide spectral range  -- from terahertz to optical
frequencies. Comparison with experiments on absorption properties
of CNT-based composites allowed proposing a qualitative
interpretation of observed peculiarity -- resonant-like behavior
of the absorbance below the range of interband transition. We
state the peak to be due to geometrical (antenna) resonance of
surface waves in finite-length CNTs. As the phase velocity $v$ of
surface wave in CNT is $50 - 100$ times smaller  than the speed of
light in vacuum, the first geometrical resonance of this wave
occurs for the tube length $50-100$ times smaller than  the
incident wavelength. In other words, for typical CNT lengths $0.1
-1\,\mu$m the antenna resonance-induced peak is shifted into
far-IR or terahertz region. Note that a hypothetical rode with
bulk conductivity of gold and with radius below $0.5$ nm is also
characterized by a strong slowing down of  plasmon-polariton mode
\cite{Crozier} ($c /v = 50 - 100$). However, fabrication of such
thin metallic rods is technologically difficult problem. Another
advantage of CNT comparing with metallic antenna is considerable
time of electron mean free path in the regime of Drude
conductivity, that likely leads to stronger field localization
near CNT, than near the realistic metallic antenna. Therefore,
CNTs look advantageous for  systems operating in far-IR or
terahertz regions, directed to both optoelectronic and biomedical
applications.

\acknowledgments

The research was partially supported by the INTAS under projects
05-1000008-7801 and 06-1000013-9225, International Bureau BMBF
(Germany) under project BLR 08/001, and the Belarus Republican
Foundation for Fundamental Research and CNRS (France) under
project F07F-013. M.V.S. acknowledges a support through the World
Federation of Scientists fellowship.


\begin{thebibliography}{99}


\bibitem{Reich_b04} S. Reich, C. Thomsen, J. Maultzsch,  Carbon Nanotubes. Basic Concepts and Physical Properties, Wiley-VCH, Berlin,
(2004)

\bibitem {Slepyan99} G. Ya. Slepyan, S. A. Maksimenko, A. Lakhtakia, O. Yevtushenko, and A. V. Gusakov,  \textit{Phys. Rev. B.} 60, 17136 (1999)

\bibitem{Maksim00} S. A. Maksimenko and G. Ya. Slepyan,
Electrodynamic properties of carbon nanotubes, in Electromagnetic
Fields in Unconventional Materials and Structures, edited by O. N.
Singh and A. Lakhtakia, Wiley, New York, (2000), pp. 217-255.

\bibitem{Hagmann_05} J. Hagmann, \textit{IEEE Trans. Nanotechnol.} 4,
289 (2005).

\bibitem{Rybczynski_07_APL} J. Rybczynski, K. Kempa, A. Herczynski, Y. Wang, M. J. Naughton,
Z. F. Ren, Z. P. Huang, D. Cai,  M. Giersig, \textit{Appl. Phys.
Lett.} 90, 021104 (2007)

\bibitem{Raychowdhury_06} A. Raychowdhury, and  K. Roy, \textit{IEEE Trans.
Computer-Aided Design} 25, 58 (2006)

\bibitem{Miano_06} G. Miano, F. Villone, \textit{IEEE Trans. Antennas Propag.}
54, 2713 (2006)

\bibitem{Chiariello_07} A. G. Chiariello
and G. Miano,  \textit{COMPEL: Int. J. for Computations and
Mathematics in Electrical and Electronic Engineering} 26, 571
(2007)

\bibitem{Maffucci_08} A. Maffucci, G. Miano, F. Villone, \textit{Int. J.
Circuit Theory and Appl.} 36, 31 (2008)

\bibitem{Li_08} H. Li, W.-Y. Yin, K. Banerjee, and J.-F. Mao, \textit{IEEE Trans.
Electron Devices} 55, 1328 (2008)

\bibitem{Wang_04_APL} Y.~Wang, K.~Kempa, B.~Kimball, G.~Benham, W.~Z. Li, T.~Kempa, J.~Rybczynski,   A.~Herczynski, and Z.~F. Ren,  \textit{Appl. Phys. Lett.} 85,  2607 (2004)

\bibitem {Hanson05} G. W. Hanson, \textit{IEEE Trans. Antennas Propag.} 53, 3426 (2005)

\bibitem {Burke06} P. J. Burke, S. Li, and Z. Yu, \textit{IEEE Trans. Nanotechnol.} 5, 314
(2006)

\bibitem {Slepyan06} G. Ya. Slepyan, M. V. Shuba, S. A. Maksimenko,
and A. Lakhtakia, \textit{Phys. Rev. B.} 73, 195416 (2006)

\bibitem {Hanson06}  J. Hao, and G. W. Hanson, \textit{Phys. Rev. B.} 74, 035119 (2006)

\bibitem{Kempa_07} K. Kempa, J. Rybczynski, Z. Huang, K. Gregorczyk, A. Vidan, B. Kimball, J. Carlson, G. Benham,
Y. Wang, A. Herczynski, Z. F. Ren,  \textit{Advanced Materials}
19, 421 (2007)
\bibitem {Shuba07}  M. V. Shuba, S. A. Maksimenko, and A. Lakhtakia, \textit{Phys. Rev. B.} 76, 155407 (2007)

\bibitem {Nemilentsau} A. M. Nemilentsau, G. Ya. Slepyan, S. A. Maksimenko,  \textit{Phys. Rev. Lett.} 99, 147403 (2007)

\bibitem{Wang_08_IJIRMW} Y. Wang, Q. Wu, W. Shi, X. He, X. Sun,  T. Gui, Int. \textit{J. Infrared
Millim. Waves}  29, 35 (2008)

\bibitem{Hanson08} G. W. Hanson, \textit{IEEE  Antennas Propag. Mag.}, to be published, (2008)

\bibitem {Maksimenko07} S. A. Maksimenko, G. Ya. Slepyan, A. M. Nemilentsau, and M. V.
Shuba, \textit{Physica E.} 40, 2360 (2008)

\bibitem{Rutherglen} C. Rutherglen, and P. Burke,
\textit{Nano Lett.}  7, 3296 (2007)

\bibitem{Iensen} K. Iensen, I. Weldon, H. Garsia, and A. Zettl,
\textit{Nano Lett.}  7, 3508 (2007)

\bibitem{Misewich_03_Sci} J. A. Misewich, R. Martel, Ph. Avouris, J. C. Tsang, S. Heinze, J. Tersoff,
\textit{Science} 300, 783 (2003)


\bibitem{Chen_05_Sci} J. Chen, V. Perebeinos, M. Freitag, J. Tsang, Q. Fu, J. Liu, and P. Avouris
\textit{Science} 310, 1171 (2005)

\bibitem{Kibis3} O. V. Kibis, M. E. Portnoi, \textit{Tech. Phys. Lett.}
31, 671 (2005)

\bibitem{Kibis_07_NL} O. V. Kibis, M. Rosenau da Costa, and M. E.
Portnoi, \textit{Nano Lett.} 7, 3414 (2007)

\bibitem{Batrakov_06_SPIE}
K. G. Batrakov, P. P. Kuzhir, S. A. Maksimenko, \textit{Proc.
SPIE} 6328, 63280Z (2006)

\bibitem{Kuzhir_07_SRIMN} P. Kuzhir, K. Batrakov, S. Maksimenko,
\textit{Synthesis and Reactivity in Inorganic, Metal-Organic and
Nano-Metal Chemistry} 37, 341 (2007)

\bibitem{Batrakov_08_PhE} K. G. Batrakov, P. P. Kuzhir, and S. A. Maksimenko, \textit{Physica E.} 40, 1065 (2008)

\bibitem{O'Connel_02} M. J. O'Connell,  S. M. Bachilo, C. B. Huffman, V. C. Moore,
M. S. Strano,  E. H. Haroz, K. L. Rialon,  P. J. Boul, W. H. Noon,
C. Kittrell, J. Ma,  R.H. Hauge, R. B. Weisman, R. E. Smalley,
 \textit{Science} 297, 593 (2002)

\bibitem {Bommeli} F. Bommeli, O. L. Degiorgi, P. Wachter, W. S. Bacsa, W. A. de Hee and
L. Forro,  \textit{Solid State Commun.} 99, 513 (1996)

\bibitem {Ugawa} A. Ugawa, A. G. Rinzler, and D. B. Tanner,  \textit{Phys. Rev. B.} 60, R11305 (1999)

\bibitem {Zhao} H. Hu, B. Zhao, M. A. Hamon, K. Kamaras, M. E. Itkis, R. C. Haddon,  \textit{J. Am. Chem. Soc.} 125, 14893 (2003)

\bibitem {Kang} C. Kang, I. H. Maeng, S. J. Oh, S. C. Lim, K. H. An, Y. H. Lee,
and J.-H. Son,  \textit{Phys. Rev. B}. 75, 085410 (2007)

\bibitem{Kam_05} N. W. S. Kam, M. O'Connell, J. A. Wisdom, and H. Dai, \emph{Proc. Natl. Acad. Sci. USA},  102,
11600 (2005)

\bibitem{Panchapakesan_05} B. Panchapakesan, S. Lu, K. Sivakumar, K. Teker, G. Cesarone, and E.
Wickstrom, \textit{NanoBiotechnology}, 1, 133 (2005)

\bibitem{Gannon_07} C. J. Gannon, P. Cherukuri, B. I Yakobson, L. Cognet, J. S Kanzius,
C. Kittrell, R B. Weisman, M. Pasquali, H. K. Schmidt, R. E.
Smalley, S. A Curley,
\textit{Cancer} 110, 2654 (2007)

\bibitem{Yang_07} W. Yang, P. Thordarson, J. J. Gooding, S. P. Ringer and F. Braet, \textit{Nanotechnology} 18,  412001
(2007)

\bibitem{Zharov_05} V. P. Zharov, E. N. Galitovskaya, C. Johnson, T. Kelly,  \textit{Lasers
Surg. Med.} 37, 219 (2005)

\bibitem{Lapotko_2007} D. O. Lapotko, E. Y. Lukianova-Hleb, A. A. Oraevsky,
\textit{Nanomedicine} 2, 241 (2007)

\bibitem{Huang_07} X. Huang, P. K. Jain, I. H. El-Sayed, and M. A. El-Sayed, \textit{Lasers
Med. Sci.} DOI: 10.1007/s10103-007-0470-x (2007)

\bibitem{Gemmer_b05} J. Gemmer, M. Michel and G. Mahler, Quantum
Thermodynamics: Emergence of Thermodynamic Behaviour within
Composite Quantum Systems, Lect. Notes Phys. 657, Springer,
Berlin-Heidelberg, (2005).

\bibitem{Volz_b07} Microscale and Nanoscale Heat Transfer, Topics Appl. Phys. 107, edited by
S. Volz, Springer, Berlin-Heidelberg, (2007).

\bibitem{Berber_00} S. Berber, Y.-K. Kwon, D. Tomanek, \emph{Phys. Rev.
Lett.} 84, 4613 \textbf{(2000)}.

\bibitem{Che_00} J. Che, T. Cagin, W.A. Goddard III, \emph{Nanotechnology}
11, 65 (2000).

\bibitem{Kuroda_05} M. A. Kuroda, A. Congellaris, J.-P. Leburton, \emph{Phys. Rev.
Lett.} 95, 266803 (2000).

\bibitem{Pop_06} E. Pop, D. Mann, Q. Wang, K. Goodson, and H. Dol,
\emph{Nanolett.} 6, 96 (2006)

\bibitem {Dresselhaus02} M. S. Dresselhaus , G. Dresselhaus , A. Jorio , A. G. Souza Filho,
R. Saito,  \textit{Carbon} 40, 2043 (2002)

\bibitem {Kane} C. L. Kane and E. J. Mele,  \textit{Phys. Rev. Lett.} 78, 1932 (1997)

\bibitem{Ilyinsky} A. S. Ilyinsky, G. Ya. Slepyan, and A. Ya. Slepyan, Propagation,
scattering and dissipation of electromagnetic waves, Peter
Peregrinus, London, (1993)

\bibitem{Mint_98} J.~W.~Mintmire, and C.~T.~White, \textit{Phys. Rev.  Lett.}  81, 2506
(1998)

\bibitem{Novotny} L. Novotny, and  B. Hecht, Principles of Nano-Optics, Cambridge University Press, Cambridge, (2006)

\bibitem {Crozier}  K. B. Crozier, A. Sundaramurthy, G. S. Kino, and C. F. Quate, \textit{J. Appl. Phys.} 94, 4632 (2003)

\bibitem {Kappeler}  R. Kappeler, D. Erni, C. Xudong, and L. Novotny, \textit{J. Comput. Theor. Nanoscience} 4, 686 (2007)

\bibitem {Krug}  J. T. Krug, E. J. Sanchez, and X. S. Xie, \textit{J. Chem. Phys.} 116, 10895 (2002)

\bibitem {Hillenbrand}  R. Hillenbrand, T. Taubner, and F. Keilmann, \textit{Nature} (London) 418, 159 (2002)

\bibitem {Puscasu}  I. Puscasu, D. Spencer, and G. D. Boreman, \textit{Appl. Opt.} 39, 1570 (2000)

\bibitem {Hilt}  O. Hilt, H. B. Brom, and M. Ahlskog, \textit{Phys. Rev. B.} 61, R5129
(2000)

\bibitem{Boyd_84} G. T. Boyd, T. Rasing, J. R. R. Leite, Y. R. Shen,
\textit{Phys. Rev. B.} {30}, 519 (1984)


\end{thebibliography}
\end{document}